\documentclass[11pt,a4paper]{article}
\usepackage[makeroom]{cancel}
\usepackage{bm}
\usepackage{amsmath}
\usepackage{amsfonts}
\usepackage{graphicx}
\usepackage{nicefrac}
\usepackage{authblk}
\usepackage{cite}
\usepackage[left=2.54cm,top=2.54cm,right=2.54cm,bottom=2.54cm,nohead]{geometry}
\DeclareGraphicsExtensions{.png,.jpg,.pdf}
\usepackage{epstopdf}
\usepackage{float}
\usepackage{subfigure}
\usepackage{fouriernc} 
\usepackage{enumitem}

\title{Theory of expansion of porous carbon electrodes \\ in aqueous solutions according to the Donnan model}

\usepackage{authblk}
\makeatletter
\renewcommand\AB@authnote[1]{\textsuperscript{\normalfont#1}}

\makeatother

\author[]{P.M. Biesheuvel}

\affil[]{Wetsus, European Centre of Excellence for Sustainable Water Technology, Leeuwarden, The Netherlands.}

\date{} 

\begin{document}


\renewcommand{\t}{\widetilde}
\renewcommand{\t}{}
\newcommand{\s}[1]{\mathrm{_{#1}}}

\maketitle

\begin{abstract}
The Donnan model describes the electric double layer structure inside carbon micropores and is an essential element of larger-scale models for capacitive porous carbon electrodes for energy storage and water desalination. The Donnan model was not yet developed to describe the expansion of carbon electrodes upon charging. Here we provide a simple theory which predicts that the one-dimensional expansion of a carbon particle scales with electrode charge density according to a $\nicefrac{4}{3}^\text{rd}$ power.

\end{abstract}

\noindent\rule{12cm}{0.4pt}

\bigskip

The Donnan model is an electric double layer model suitable for the physical description of porous carbon electrodes for energy storage and water treatment purposes, such as for water desalination by capacitive deionization~\cite{Muller_Kastening}. Using only a few analytical expressions, it describes the equilibrium adsorption of salt ions into carbon micropores as function of applied voltage, external salinity, micropore volume, and chemical surface charge. The Donnan model makes use of the fact that in carbon micropores the diffuse layers are strongly overlapped and a common value for electric potential can be assumed across the pore. 


Over the past years, the Donnan model has been successfully used to describe data for salt adsorption in capacitive deionization (CDI) from single salt solutions and salt mixtures~\cite{Biesheuvel_2015}. Especially important is the recent discovery of the role of chemical ``immobile'' surface charge, which modulates salt adsorption~\cite{Suss_Arxiv,Suss_Colcom,Gao_WR}. This chemical charge can be synthesized into the material, or will develop when the electrode is in use in water which contains oxygen gas. The presence of chemical surface charge leads to various novel modes of CDI-operation such as inverted CDI, enhanced CDI, and extended-voltage CDI~\cite{Suss_Colcom}. Faradaic processes in porous electrodes depend on the EDL-structure as well as on the concentration of reactive ions in the micropores, which are parameters that are predicted by the Donnan approach. Thus, the Donnan model makes it possible to self-consistently describe Faradaic processes in porous electrodes~\cite{RJE}. 


The Donnan model can be used as a theoretical framework in its own right, but it can also be treated in a more utilitarian manner, simply as a suitable mathematical structure to collect experimental data of the EDL-structure (e.g., salt adsorption and charge vs. voltage), or data from more elaborate DFT calculations or MC simulations. The Donnan model then simply serves as a translation tool between output of detailed EDL-structure calculations, and macroscopic transport modeling. This is of relevance because to study processes involving carbon electrodes, it is essential to set up transport theory for ion electrosorption and charge transfer on the scale of the entire porous electrode film and cell device. Because it is not possible to directly combine results of simulations with transport theory, the Donnan modeling framework can be of use to summarize all data from advanced simulations. 

\medskip

\noindent Three further remarks on the Donnan approach are as follows.
\begin{itemize}[leftmargin=*]
\item{First, the Donnan model is typically used with the assumption of fast micropore $\longleftrightarrow$ macropore ion exchange, i.e., the content of the micropore is at equilibrium with the salt solution just outside the pore (which is the macropore region, which provides ion transport pathways across the electrode). However, this assumption can be relaxed in a model that includes a micropore-to-macropore mass transfer resistance.} 
\item{Second, the author of this document made significant use in the past of Donnan models that use a ``micropore attraction term $\mu\s{att}$'', either as a constant or as a parameter that is inversely related to pore ions concentration. This modification describes some data very well, but the extra term also poses questions, e.g., how it depends on the ion type and ion mixtures. This model also does not describe the new CDI-modes that depend on chemical surface charge. Therefore, the author presently advocates Donnan models without the $\mu\s{att}$-term but including chemical surface charge.} 
\item{Third, to describe data quantitatively, accurate Donnan models require two types of micropore regions, each with its own electrostatic environment and chemical charge, where typically one region contains acidic surface groups (from carboxylic acid) and the other basic groups~\cite{Biesheuvel_2015,Gao_WR}. The latter groups can be synthesized into the pores with amine-chemistry, but are also present because the graphite basal planes have an affinity to adsorb protons which gives the pore walls a net positive charge.}
\end{itemize} 


Up until now, the Donnan model has not yet been used to describe how carbon particles expand upon charging~\cite{Soffer_Folman,Golub_JEC,Golub_carbon}. It was recently argued that the Donnan approach ``falls short ... to quantify electrosorption-induced electrode swelling''~\cite{Prehal_NE}. That may well be true, but momentarily cannot be ascertained, because the Donnan model was not yet developed to describe carbon expansion (electrode swelling). To make it possible to test the Donnan model on this point, we here present equations for carbon particle expansion vs. electrode charge according to the Donnan model. Future experimental work and simulations can test the scaling relation that will be put forward. 


In this paper we set up a simple Donnan model to describe the expansion of porous carbons when they are charged. The Donnan model takes the simple approach of describing ions as simple point charges (though it can be extended, e.g., by including a pore-modified Carnahan-Starling equation of state which includes ion volume), and makes use of a common pore ``Donnan'' potential. Ion concentration in the micropore is simply related to salt concentration outside the pore by the Boltzmann-relation. Additional features are chemical surface charge and the Stern capacitance, which relates electronic charge in the carbon to Stern potential. The origin for the Stern capacitance can be the non-zero closest-approach distance of ions to the pore wall, or can also relate to space-charge effects within the carbon itself.


To relate pressures to expansion we must choose a specific geometrical shape of the micropore, for which we choose a hollow cylinder. Alternatives are a spherical geometry but this seems unrealistic because a closed sphere does not allow ions to enter, or alternatively, a planar geometry, but here expansion is infinite at any pressure difference. Thus we use the cylindrical geometry. In the thin-shell, small-expansion, limit, the radial expansion of a cylinder relates to pressure difference across the pore wall according to
\begin{equation}
\frac{\Delta R}{R_0}=\frac{1}{E}\frac{R_0}{\delta}\Delta P
\end{equation}
where $R_0$ is the pore radius, $\Delta R$ the change in radius upon applying a pressure difference $\Delta P$, $E$ Young's modulus of the carbon, and $\delta$ the thickness of the carbon walls. For a material composed of cylinders with all possible orientations, the linear relative expansion of the material (assuming it is unconstrained) is given by
\begin{equation}
\frac{\Delta \ell}{\ell}=\left({\frac{\Delta R}{R_0}}\right)^{\frac{2}{3}}
\end{equation}
while the volumetric expansion $\Delta V / V$ scales with $\Delta \ell /\ell$ to the power 3.


The pressure difference between inside and outside of the pore, $\Delta P$, has two components. The first contribution is from the osmotic pressure difference, and is given by
\begin{equation}
\Delta P\s{osm}=RT \> \left(c\s{ions,mi}-c\s{ions,\infty}\right)
\end{equation}
where $c\s{ions,mi}$ is the total ions concentration in the carbon micropores, and $c\s{ions,\infty}$ is the ions concentration outside the pore. 
For a 1:1 salt solution, the internal micropore ions concentration relates to the volumetric ionic charge density by
\begin{equation}
c_\text{ions,mi}^2=\left(\sigma/F\right)^2+4\cdot c_{\infty}^2
\end{equation}
where $c_\infty$ is the salt concentration outside the pore. 


The pressure due to the field energy in the Stern layer follows from an electrical energy analysis, which results in
\begin{equation}
\Delta P\s{fe}=\frac{\sigma^2}{4C\s{S}}
\end{equation}
where in general $\sigma$ is the electronic charge density residing in the carbon (defined per unit micropore volume). Assuming zero chemical charge, the electronic charge density in the carbon is equal to the ionic charge density. The Stern potential is defined per unit pore volume (unit F/m$^3$). In line with the Donnan approach to neglect potential gradients in the water-filled interior of the micropore, we only consider the field energy related to the Stern capacitance, and not in the aqueous phase.
 

Thus, according to the Donnan model, we arrive for the linear expansion of a macroscopic carbon particle that has no chemical surface charge groups, in a 1:1 salt, at

\begin{equation}
\frac{\Delta \ell}{\ell} \propto \left\lbrace 2RTc_\infty \> \left\lbrace\sqrt{\left(\frac{\sigma}{2Fc_\infty}\right)^2+1}-1\right\rbrace + \frac{\sigma^2}{4 C\s{S}} \right\rbrace ^{\scalebox{1.3}{$\nicefrac{2}{3}$}}
\end{equation}
which in the limit of low charge simplifies to
\begin{equation}
\frac{\Delta \ell}{\ell} \propto \left\lbrace \frac{RT}{4 F^2 c_\infty} + \frac{1}{4 C\s{S}} \right\rbrace \vphantom{\Sigma^2}^{\scalebox{1.2}{\nicefrac{2}{3}}}  \cdot \sigma \vphantom{\Sigma^2}^{\scalebox{1.2}{\nicefrac{4}{3}}}
\end{equation}
and in the limit of high charge becomes
\begin{equation}
\frac{\Delta \ell}{\ell} \propto \left\lbrace  \frac{1}{4 C\s{S}} \right\rbrace \vphantom{\Sigma^2}^{\scalebox{1.2}{\nicefrac{2}{3}}}  \cdot \sigma \vphantom{\Sigma^2}^{\scalebox{1.2}{\nicefrac{4}{3}}}.
\end{equation}

Thus, except for a region of intermediate charge, the linear expansion of a carbon particle is expected to scale with charge density to the power \nicefrac{4}{3}. At low charge, a carbon particle will expand when the salt concentration goes down (and charge stays the same), while at high charge there is no dependence of expansion on salt concentration.

\end{document}